# Preparing HPC Applications for the Exascale Era: A Decoupling Strategy

Ivy Bo Peng*, Roberto Gioiosa†, Gokcen Kestor†, Erwin Laure* and Stefano Markidis*
*School of Computer Science and Communication, KTH Royal Institute of Technology, Sweden
†Computational Science and Mathematics Division, Pacific Northwest National Laboratory, USA
‡Email: ivybopeng@kth.se

*Abstract*— Production-quality parallel applications are often a mixture of diverse operations, such as computation- and communication-intensive, regular and irregular, tightly coupled and loosely linked operations. In conventional construction of parallel applications, each process performs all the operations, which might result inefficient and seriously limit scalability, especially at large scale. We propose a *decoupling* strategy to improve the scalability of applications running on large-scale systems. Our strategy separates application operations onto groups of processes and enables a dataflow processing paradigm among the groups. This mechanism is effective in reducing the impact of load imbalance and increases the parallel efficiency by pipelining multiple operations. We provide a proof-of-concept implementation using MPI, the de-facto programming system on current supercomputers. We demonstrate the effectiveness of this strategy by decoupling the reduce, particle communication, halo exchange and I/O operations in a set of scientific and data-analytics applications. A performance evaluation on 8,192 processes of a Cray XC40 supercomputer shows that the proposed approach can achieve up to 4× performance improvement.

*Keywords*-HPC Streaming Computing, Decoupling, Exascale Computing, MPI Data Streams

## I. INTRODUCTION

Large-scale HPC applications already face serious challenges in achieving high parallel efficiency on current petascale supercomputers. These challenges will further increase on the exascale systems that are expected to provide billion-way parallelism. Improving scalability in such scenarios becomes a daunting task for many factors, including the increased cost of communication and data movement, power and energy constraints, high fault rates, and load imbalance [1], [2]. For instance, the collective operations will significantly impact scalability because their complexities increase with the number of processes. More complicated cases are the real-world applications consisting of multiple operations that are characteristically different in terms of scalability. For example, molecular dynamics codes perform computation-intensive operations on particle-particle interactions as well as communication-intensive fast Fourier transform (FFT). Typically, parallel applications performs all these operations on each process. In this work, we explore whether higher level of parallelism can be achieved by separating the different operations onto different processes and pipelining them to progress in parallel. This concept is similar to the instruction-level parallelism implemented on superscalar processors, where different execution units concurrently work on independent data items.

Parallel applications often consist of distinct stages of operations. Data exchange and synchronization occur at the completion of an operation. This model assumes equal work to result in equal run time on each process. However, on large-scale machines, the interference from system noises is unavoidable [3] and it will only become worse on extreme-scale systems. Other external factors, such as higher temperature variance, also vary the speed of processors. Moreover, it is common that scientific applications (e.g., unstructured meshes or adaptive grids) allocate variable amount of workload to processes. For imbalanced applications, the cost of data exchange includes both the data transfer time and the idle time that a process waits for its delayed peer [4], [5].

We propose a decoupling strategy to separate operations onto groups of processes and to enable a dataflow processing paradigm among these groups. When operations are decoupled, the programmers can apply application-specific knowledge to further optimize the operation. For instance, they can use data aggregation scheme and data transfer scheduling on communication-intensive operations. On the other hand, as different groups of processes perform different operations, multiple operations can progress in parallel in a pipelined fashion. The dataflow among groups allows each group to perform on the first available data without waiting for a particular communication peer. This helps reduce the impact from the process imbalance in applications. Figure 1 illustrates an example of decoupling computation, communication, I/O, and data analytics operations within one application to be performed by disjoint groups.

On the way to exascale computing, the programmers can employ either incremental or disruptive approaches to improve the scalability of their applications. Incremental approaches use high-performance implementation in the traditional programming system, such as RMA operations in MPI [6]. This approach can reduce the idle time on imbalanced processes but is still limited in pipelining operations because they are coupled on all processes. Disruptive approaches port applications to programming languages that provide load balancing and dynamic execution, e.g., Cilk++, Charm++ and Chapel [7], [8], [9]. This approach requires significant development efforts and a long implementation time for large production-quality applications. In this work, we explore a strategy that is an intermediate of the two. With exascale computing expected in 2020-2024, we propose a solution that requires incremental changes to the existing applications and still provides considerable performance improvement and increased flexibility to cope with external factors.

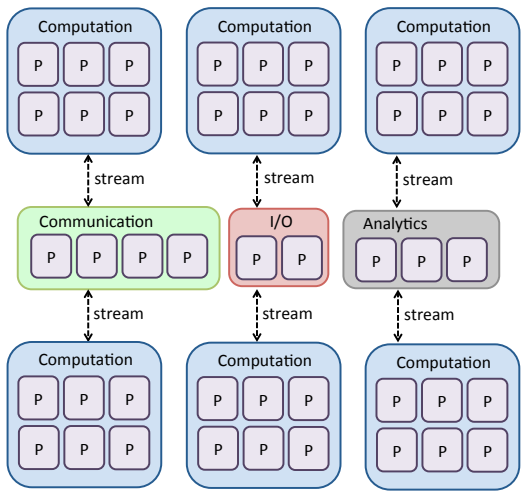

Fig. 1: Decouple computation, communication, I/O and analytics operations in one application. The different groups are linked with data streams.

To evaluate the feasibility of our decoupling strategy, we provide a proof-of-concept implementation atop MPI [10], which is the dominant programming standard on current supercomputers. We show that, by pipelining operations on different groups and optimizing the decoupled operations, applications running on large-scale systems can have an improved scalability compared to the conventional model and the performance gap widens as the number of processes increases. For instance, a decoupled implementation of the MapReduce application results in a $4\times$ performance improvement compared to the reference implementation on 8,192 MPI processes on a Cray XC40 machine. Since the performance gap increases at scale, we expect that the performance benefits from the proposed strategy will also widens on future exascale systems.

Overall, we made the following contributions in this work:

- We propose a novel decoupling strategy to separate operations onto groups of processes at application-level. Our strategy establishes a dataflow processing paradigm among groups.
- We analyze the performance of the proposed strategy, identify the key factors and provide a guideline to select operations that can benefit from decoupling at large-scale.
- We provide an implementation of the decoupling strategy in an MPI-based stream library.
- We evaluate our strategy and its implementation by decoupling reduce, particle communication, halo exchange and I/O operations in a set of data-analytics and scientific applications running on a petascale testbed.

The paper is organized as follows. Sections II and III introduce the decoupling strategy and its realization in existing applications. Section IV presents case studies in real-world applications and their evaluation on petascale machines. Section V discusses related works. Section VI concludes the work.

## II. THE DECOUPLING APPROACH

In this section, we first present a real-world application as an example of motivation. We then provide an overview of the decoupling approach and a detailed analysis of the key factors of the performance. Lastly, we identify five categories of operations that are suitable for the proposed approach.

### A. Motivation

Parallel applications often consist of multiple distinct stages. Each stage mainly performs one type of operation, such as computation, communication, I/O operations, data analytics and load balancing. These phases are tightly coupled on all processes and one phase only starts after the previous phase finishes. The top panel of Figure 2 shows the execution trace of a plasma simulation application (details in Section IV) collected by HPCToolkit [11] on seven MPI processes. The trace shows that two operations, i.e. particle computation (in grey) and particle communication (in blue), are executed sequentially on all processes. The decoupling strategy separates the two operations onto two groups: $G_0$ of P1-P6 and $G_1$ of P0. After the decoupling, group $G_0$ computes particle trajectory (in grey) and group $G_1$ communicates exiting particles (in blue). The bottom panel of Figure 2 shows the execution trace of the decoupled implementation. It is clear that the decoupled implementation reduced the execution time of the application. Also, the two operations progress concurrently, overlapping on the timeline.

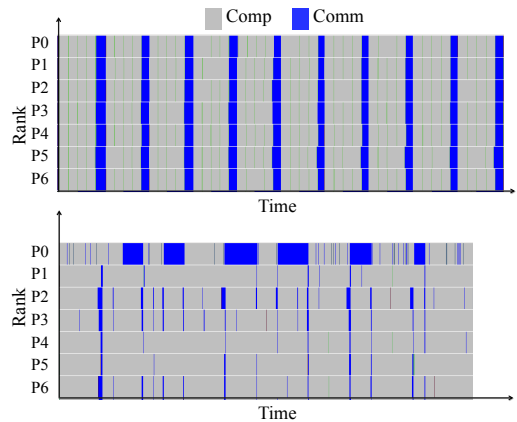

Fig. 2: HPCToolkit [11] trace of the plasma simulation code iPIC3D using reference implementation (top) and decoupled implementation (bottom).

### B. Approach Overview

We propose a decoupling strategy that does not assume a rigid and staged execution of the operations in an application. Instead, multiple operations can progress concurrently on

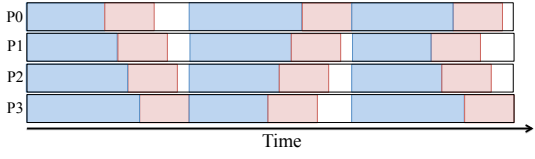

(a) The conventional model performs all operations sequentially on all processes.

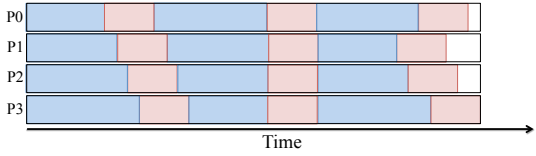

(b) Using non-blocking operation absorbs process imbalance but does not pipeline different operations.

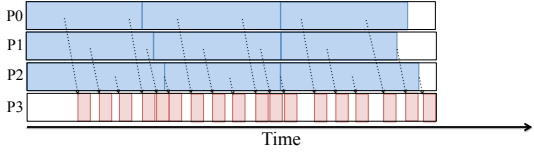

(c) The decoupling approach pipelines operations, reduces the complexity and absorbs process imbalance.

Fig. 3: The comparison of different approaches for parallel applications.

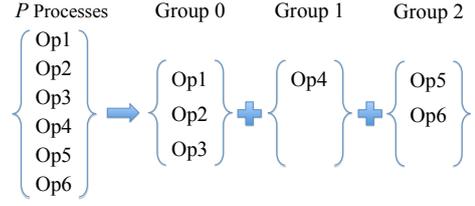

Fig. 4: Adapting an application to the decoupling model.

different groups of processes. Fine-grained, asynchronous data streams among groups effectively establish a dataflow processing paradigm. Process imbalance is absorbed by processing the first available data stream from a group instead of waiting for a particular communication peer. The complexity of decoupled operations is reduced when moving from a large number of processes to a small subset of processes. The programmers can further optimize the decoupled operations with application-specific knowledge.

We illustrate the differences between the conventional model, the non-blocking operations, and the decoupled approach in Figure 3. In the conventional model, each process performs the two operations (in red and blue, respectively). Processes are in idle state when waiting for delayed process to complete their portion of the work. In the ideal case, using non-blocking operations can reduce the idle time on imbalanced processes. Still the two operations are coupled on all processes without pipelining. Decoupling the operation (in red) to P3, reduces its complexity and enables two operations to progress concurrently.

### C. Application Adaption

Assume an application has $N$ operations, i.e., Op1 - Op$N$. This application runs on a total of $P$ processes. The first step of adopting the decoupling strategy is to form a number of processes groups. Each group $G_i$ consists of $P_i$ processes. After the group setup, the $N$ operations are mapped to the $G$ groups. The basic criteria for this operation-to-group mapping is to have all operations being mapped to exactly one group so that each group works on a subset of operations. An example of this mapping is shown in Figure 4. Originally, each of the $P$ processes needs to perform six operations. We create three groups of processes and assign them a subset of the six operations, which increases the level of parallelism with pipelining.

Next, we establish an asynchronous, fine-grained data flow among groups. Traditionally, when communication has specific commutation parties, delays on one party will directly impact the other party, resulting in idle time on the waitng process. However, when data flows from $P_i$ ($> 1$) processes in a group, the probability that all $P_i$ processes are delayed is much lower. As a result, there will always be some data available from some processes in group $G_i$. Despite the fact that process imbalance can still exist in group $G_i$, its impact to the other groups can be partially or fully eliminated.

To further improve the utilization of the network and increase the communication-communication overlapping, we use fine-grained data streams as the basic unit to loosely connect groups of processes. In this way, data exchange can occur as soon as data of a certain granularity (stream element) is ready. In comparison, the staged construction of applications creates bursty communication to the network because all processes start exchanging data at approximately the same time (the completion of an operation) and leaves the network idle during the execution.

### D. Performance Model

In order to analyze the performance of an application using the decoupling strategy, we introduce a performance model that focuses on the pipelining, imbalance absorption, complexity and overhead of the decoupled operations. To simplify the explanation without loss of generality, we assume there are two different operations Op0 and Op1 in an application. The application allocates $W_0$ and $W_1$ workload to each process for the two operations respectively. When running the application on $P$ processes, the total execution time of the conventional model is simply the sum of the two workload and the expected time of process imbalance $T_\sigma$ as presented in Eq. 1.

$$T_c = T_{W0} + T_\sigma + T_{W1}. \qquad (1)$$

Assume Op1 is decoupled to a subset of $\alpha \cdot P$ processes ($0 \leq \alpha \leq 1$). There are only $(1-\alpha) \cdot P$ processes left to carry out Op0 so that each process will have proportionally more workload, i.e. $\frac{1}{(1-\alpha)} \cdot W_0$. Similarly, the workload for Op1 is only performed by $\alpha \cdot P$ processes so that each process has $\frac{1}{\alpha} \cdot W_1$ workload. Since the complexity of Op1 is changed when performed by a subset of processes, we use $T'_{W1}$ to denote the updated time of Op1. Because these two operations progress in parallel, the total execution time equals to the longer one of the two as indicated by Eq. 2.

$$T_d = \max\{\frac{1}{(1-\alpha)} \cdot T_{W0} + T_\sigma, \frac{1}{\alpha} \cdot T'_{W1}\} \quad (2)$$

Define the effective pipeline between Op0 and Op1 as the fraction of Op0 that is overlapped with Op1. Let $\beta$ denote the portion of Op0 without overlapping. For instance, $\beta = 0.3$ indicates that when Op0 has completed by 30%, Op1 starts progressing. We apply an pessimistic assumption that Op1 will always finish after Op0 as in the conventional model. In this way, the total execution time can be rewritten as in Eq. 3. This equation shows that in the worst case of no pipeline ($\beta = 1$), the execution time equals to the sum of the two operations. In the best case of perfect pipeline ($\beta = 0$), the execution time only equals to the decoupled operation.

$$T_d = \beta \cdot [\frac{1}{(1-\alpha)} \cdot T_{W0} + T_\sigma] + \frac{1}{\alpha} \cdot T'_{W1} \quad (3)$$

Note that the data flow among groups adds overhead for decoupling operations. The overhead comes from the construction of data elements and from calling the function for injecting the data streams. Let $o$ be the overhead of each stream element. Assume that a total of $D$ data is transferred between the two groups and that the data stream has a granularity of $S$. In total, decoupling Op1 imposes $\frac{D}{S} \cdot o$ overhead. Also note that $\beta$ is a function of $S$ as the finer grain the stream element is, the higher pipelining can be achieved. Taking the overhead of decoupling into consideration, the execution time is refactored as in Eq. 4.

$$T_d = \beta(S) \cdot [\frac{1}{(1-\alpha)} \cdot T_{W0} + T_\sigma + \frac{D}{S} \cdot o] + \frac{1}{\alpha} \cdot T_{W1}' \quad (4)$$

It is clear from the above equation that there are two main performance factors for the decoupling approach. First, the granularity of data streams should balance the time for pipelining and overhead. Data streams of fine granularity can improve the pipelining of operations and the absorption of process imbalance but at a cost of higher overhead. It is also important to note that some networks are more prone to congestion when a large number of messages are in-flight. On the other hand, coarse-grained data streams reduce the overhead but also limit the efficiency of hiding process imbalance. Second, the expected time of the decoupled operation ($T'_{W1}$) determines whether an operation is suitable to be decoupled. If the decoupled operations can be aggressively optimized on the decoupled group, either with application-specific knowledge or because its complexity decreases when moving to a smaller number of processes, they are good candidates for decoupling.

Memory consumption in the decoupling approach is less or equal to the one in the conventional model. When a data stream reaches its destination, it is either processed and discarded, or buffered. The former case is the most common scenario and the memory consumption is bound to the granularity of data streams $S$, where $S << D$. In the latter case, the memory consumption equals to the total amount of transferred data $D$.

Overall, the decoupling strategy has an efficient memory cost compared to the conventional model.

*E. Identify Operations for Decoupling*

We summarize five categories of operations that are suitable for decoupling:

- **Orthogonal operations with little data dependency**. When operations can be independently performed on different groups, no data flow is needed, the decoupling model benefits from pipelining these operations working on separate data. This use case is very similar to the superscalar processor, where execution units concurrently operate on independent data items.
- **Operations with high complexities on large numbers of processes**. The complexity can come from various aspects, such as communication, memory consumption or the consistency semantics. For instance, collective operations that are difficult to optimize at large scale can be executed on a subset of processes.
- **Operations with large execution time variance**. The decoupling approach is designed to reduce process imbalance by using fine-grained data flow among groups of processes. When operations have variable execution time, the data flow is more evenly distributed over the execution, improving the communication-communication overlapping within the application.
- **Operations that continuously generate data flow throughout the execution**. Conventional applications only start data exchange at the completion of a stage, this manually creates bursty communication in the network while the network is idling before the completion of the stage. The decoupling approach evenly distribute the communication over the execution and thus improves the utilization of the network.
- **Operations that can benefit from special-purpose computing unit** (i.e., large memory compute nodes, nodes equipped with burst buffer, and special I/O nodes). The next-generation supercomputers are likely to feature heterogeneous computing units. Different operations might benefit from different hardware, decoupling them can be one way to exploit the advantage of hybrid platform.

Specific optimizations can be applied to the decoupled operations when they are assigned to a dedicated group. For instance, when communication-intensive operations are decoupled, optimizations that aggregate data and schedule the data transfer are often used. When I/O-intensive operations are decoupled, the dedicated group can use aggressive buffering optimization to reduce the interaction with the file systems. When data-analytics operations are decoupled, the dedicated group can call an independent data-analytics application to process the data without interfering with the remaining processes. This optimizations can be conveniently defined using the interface provided in the MPIStream library described in the next section.

## III. THE IMPLEMENTATION

In this section, we introduce the implementation of the decoupling approach in an MPI-based stream library, called *MPIStream* [12]. We choose this implementation approach because MPI is currently the de-facto programming system on supercomputers and is likely to be available on exascale systems [1]. Alternatives for implementing the decoupling model in applications are discussed in Section V. When using the library to decouple an operation, the application remains unchanged in large part. The incremental modifications to the application require much less efforts than refactoring the application to another programming models or languages. In the remainder of this section, we introduce the key functions of the library and demonstrate their usage through an example.

### A. MPIStream Library

The MPIStream library is implemented in $C$ and developed atop of MPI [10] persistent communication. It uses similar naming conventions for functions as in MPI. The common terms in the library are explained here. A *stream* is an asynchronous data flow from the data source to the the destination over a communication *channel*. As the data flow has a direction, the group from which the data streams are originated is called *data producer* and the group to which the data reaches is called *data consumer*. The basic unit of a stream is called *stream element*. Stream elements are usually small in size and are injected into the channel as soon as data for one stream element is ready. This mechanism supports the asynchronous and fine-grained data flows between groups. The MPIStream library provides interfaces for the programmer to define the *operator* attached to a data stream for processing stream elements.

We hereby list the major functions used to implement the decoupling strategy in applications:

1) **Setup groups of data flow**. The first step is to establish a communication channel between the groups of processes using the `MPIStream_CreateChannel()` function. The optimal ratio between groups ($\alpha$ in Eq. 4) is application-specific and depends on the specific values of the workload for each operation, the total amount of transferred data and the distribution of data flow through the execution in an application.
2) **Define the stream element**. This step defines the granularity of the data flow ($S$ in Eq. 4). As discussed in Section II, the granularity of the stream element determines the efficiency of pipelining different operations and the overhead of the data flow. The structure of the stream element is defined in an MPI (derived) datatype to support non-contiguous memory layout and zero-copy streaming.
3) **Define the operator attached to the data stream**. The stream elements are processed on the first-come-first-served basis to enable absorption of process imbalance. The operator defined on the data stream is applied on-the-fly to the arrived stream elements. This is defined through the `MPIStream_Attach()` function.
4) **Start each group to work on their allocated operations**. When the data producer has data for one stream element, it calls the `MPIStream_Isend()` function to asynchronously inject stream elements into the communication channel. The data consumer calls the `MPIStream_Operate()` function to process the arrived data streams.
5) **Terminate the data flow and free the communication channel at the end of the execution**. the data producer calls the `MPIStream_Terminate()` function to indicate the end of a data stream and both groups call the `MPIStream_FreeChannel()` function to release the channel.

As explained in the performance model, the effective configuration of a decoupling approach depends on several factors, including the fraction of processes dedicated to each operation, the characteristics of the application, the amount of transferred data, the frequency of data flow, and also the hardware and software stack of the underlying platform ($\alpha, S, D, o$ in Eq 4). The optimal setup could change dynamically at runtime. Currently, the library only supports static configuration of these values. An extension to support adaptive changes of the configuration is subject of a current work.

### B. Adopting The Decoupling Approach in Applications

We demonstrate the use of the library in a simple example in Listing 1. In this example, the application has two operations: calculation (`Calculation()`) and analysis of the workload distribution on processes (`analyze_workload()`). The second operation is common in the load balancing stage in applications [13]. The application keeps carrying out these two operations until it meets the termination criteria. In a conventional implementation, each process will perform these two operations. In the decoupled implementation, we assign a majority of processes to perform `Calculation()` and a small subset of processes to analyze the workload distribution.

Listing 1: A Simple Example of Decoupled Analysis
```
#include "MPIStream.h"
// a simple operation
void analyze_workload(Workload *in){
// find the distribution of workload on all processes
  min_max_median(in);
}
int main(int argc, char** argv){
  //Initialize MPI
  ...
  // establish a communication channel
  MPIStream_Channel channel;
  MPIStream_CreateChannel(is_data_producer,
    is_data_consumer,MPI_COMM_WORLD,&channel);
  // define the stream element
  MPI_Datatype streamDatatype;
  ...
  // define the decoupled operation
  MPIStream_Operation operation(&analyze_workload);
  // attach a data stream to the channel
  MPIStream stream;
  MPIStream_Attach(streamDatatype,operation,
                   &stream,&channel);
  //both groups start progressing
  if(is_data_producer){
    //start computation
    while(!done){
      Calculation(&data);
```

```
28          if( hasWorkloadChanges )
                MPIStream_Isend(&workload,&stream);
29      }
30      //terminate a data stream
31      MPIStream_Terminate(&stream);
32    }else if(is_data_consumer){
33      //start the decoupled operation
34      MPIStream_Operate(&stream);
35    }
36    // Finalize
37    MPIStream_FreeChannel(&channel);
38    // Finalize MPI
39    ....
40    return 0;
41 }
```

The first step of decoupling operations is to setup the groups of processes and then map subsets of operations to each group, indicated in line 9-12. The `MPIStream_CreateChannel()` function establishes a communication channel among the different groups. The direction of the data flow among groups is indicated through the `is_data_producer` and `is_data_consumer` parameters. The decoupled operation is attached to the data stream by the `MPIStream_Attach()` function. In the example, the decoupled `analyze_workload()` function calculates the min, max and median workload on all processes that perform `Calculation()`. Note that finding these three simple values will require three MPI reduction operations, which are often the bottleneck of scalability in large-scale applications. After the setup, the computation group and the analysis group exhibit different execution paths. Computation group only performs computation and if there are changes in their workload, they stream the information to the analysis group using the `MPIStream_Isend()` function. At the same time, the analysis group continuously process the analysis. After the computation group has indicated the end of a data flow using `MPIStream_Terminate()`, both groups release the resources by the `MPIStream_FreeChannel()` function.

## IV. EVALUATION

In this section, we evaluate the proposed approach by decoupling operations in MapReduce, Conjugate Gradient solver and a production-quality plasma simulation code iPIC3D. For each case, we describe the changes in the application and the performance results on the testbed. We show that decoupling and optimizing a suitable operation can significant improve the scalability of applications at scale.

### A. Experimental Setup

We perform the evaluation on the Beskow supercomputer, a Cray XC40 machine. The testbed has a total of 1,676 compute nodes connected with Cray Aries interconnect network. Each compute node is equipped with two Intel Xeon E5-2698v3 16-core (2.30 GHz) Haswell processors. The testbed uses the Cray Linux operating system. Applications were compiled with the Cray C compiler version 5.2.40 with optimization flag -O3 and the Cray MPICH2 library version 7.0.4.

To ensure fair comparison, we use the same total number of processes and the same total workload for the decoupling approach and the reference implementations. This results in fewer processes performing computation in the decoupling model than in the reference implementations. We perform ten runs for each application implementation and present the average and standard deviation of their execution time.

### B. MapReduce

In this evaluation, we show that when operations are irregular and can be effectively pipelined, the decoupling strategy brings the most advantage. MapReduce is a commonly used programming paradigm for processing large data sets in data-analytics applications [14]. MapReduce applications consist of a *map* and a *reduce* operation. The map operation takes an input and produces an intermediate output in key-value pair. The reduce operation aggregates the intermediate outputs into a final result. In this case study, we use an application that extracts word histogram in a large number of log files. The map operation takes each word $w$ in files and output a $(w, 1)$ pair. The reduce operation sums up all the intermediate outputs with the same key to obtain the total count for each distinct word.

A previous work has verified the feasibility of using MPI to implement scalable MapReduce applications [15]. In the reference MPI implementation, each process performs both the map and reduce operations. When all processes complete the map operation, they use the `MPI_Iallgatherv` operation to build up the global set of keys. After that, processes use the `MPI_Ireduce` operation to aggregate their intermediate outputs. It was pointed out that MPI lacks reduction operations that work on variable-sized input and output [15].

Natural language has irregular distribution of words so that the application will produce variable amount of results on processes. As the map operation continuously generates intermediate outputs, data flow is evenly distributed throughout the execution. In this way, the map and reduce operations can be pipelined to progress in parallel. Moreover, the complexity of the reduce operation naturally decreases when moving from a large number of processes to a smaller subset of processes. All these characteristics satisfy the criteria of suitable operations for decoupling in Section II-E.

We decoupled the map and reduce operations onto two groups. We link these two groups with MPI streams. The reduce group was further decoupled into one group that reduces the streams locally and one master process that aggregates the global results. The map group reads in the data sets from log files, performs the word-by-word mapping and streams out the intermediate results to the reduce group. Concurrently, the reduce group keeps processing the data streams on-the-fly. In this application, we use a large number of processes for the map operation as they need to handle the slow I/O operations. A small subset of processes (small $\alpha$ in Eq. 4) are dedicated to the reduce operation as this operation has reduced complexity on a small number of processes, i.e., $T_{W_1}' << T_{W_1}$ when $P_1 << P$.

We conduct a weak scaling test of the reference and the decoupled implementations on the testbed. We use the publicly available datasets from the Wikipedia web logs [16]. A total of 2.9 TB data is computed on 8,192 processes. Each log file has a different size, ranging from 256 MB to 1 GB. We varied the fraction of processes for the decoupled operation, indicated in different values of $\alpha$. $\alpha = 12.5\%, 6.25\%, 3.125\%$ indicates that one out of every 8, 16 and 32 processes, respectively, were used to perform the reduce operation. Figure 5 presents the results.

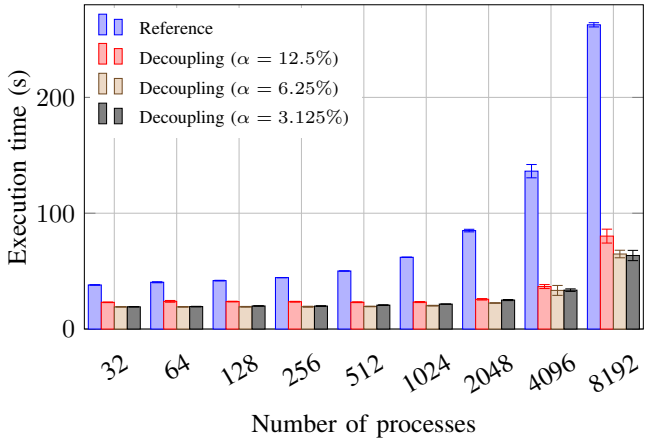

Fig. 5: Weak scaling test of the MapReduce application that computed 2.9 TB data on 8,192 processes. Different fraction of processes ($\alpha$) were used for the decoupled reduce operation.

The experiment results show that the decoupled implementation outperforms the reference implementation at scale and the improvement increases as the number of processes increases. From a $2\times$ improvement on 32 processes, the improvement increases to $4\times$ on 8,192 processes. The decoupled implementation presents almost-perfect weak scaling up to 2,000 processes. To determine the optimal value for $\alpha$, we conduct separate tests varying the values of $\alpha$. Figure 5 shows that the highest performance is achieved when one out of 16 processes are used for the decoupled operation. Note that we did not apply data aggregation to optimize the data flow within the reduce group. This could create congestion on the master process when the size of reduce group is large. For this reason, we observed an increased execution time of the decoupled implementation on 4,096 and 8,192 processes. This implies that there is more space for performance improvement within the decoupled implementation.

### C. Conjugate Gradient Solver

In this evaluation, we show that the overhead of dataflow in the decoupled implementation is no larger than the overhead of non-blocking MPI operations, even on regular operations with tight data dependency. We use the Conjugate Gradient (CG) solver that solves the Poisson equation discretized on a Cartesian uniform grid. CG solver is a widely-used iterative linear solver for symmetric positive-definitive systems. One common use case is to solve the linear system from the discretized Poisson equation. A broad spectrum of scientific applications use the CG solver as one of their key components.

Parallel implementations of the CG solver decompose the grid onto all processes so that each process only calculates the unknown variables defined in their subdomains. There are three major operations at each iteration. First, processes perform *halo exchange* to exchange with their neighbours the values on their subdomain boundaries. Second, each process carries out a discretized Laplacian operation on their local grid. Finally, all processes calculate their local residuals and then aggregate their local values to the global residual. This value is used in the next iteration of the solver. The CG solver terminates when the global residual converges below a threshold value.

The reference implementation is an open-source code[2] that uses MPI blocking and non-blocking all-to-all collective operations to implement the halo exchange [17]. [17] demonstrates the effectiveness of using non-blocking collective operations to overlap the communication and computation in the halo-exchange. We decouple the halo exchange operation onto a separate group of processes, denoted as group $G_1$. The remaining processes form the group $G_0$. At each iteration, processes in group $G_0$ stream out their boundary values and then perform the discretized Laplacian operation. At this time, they only compute the inner part of their subdomain because the boundary values are not ready. Concurrently, the processes in group $G_1$ keep on processing the arrived data streams. In group $G_0$, each process requires boundary values from six neighbour processes. Instead of communicating with six processes, the group $G_1$ aggregates these boundary values for group $G_0$ and stream them back. After group $G_0$ finishes the discretized Laplacian operation, they use the boundary values from group $G_1$ to complete the computation.

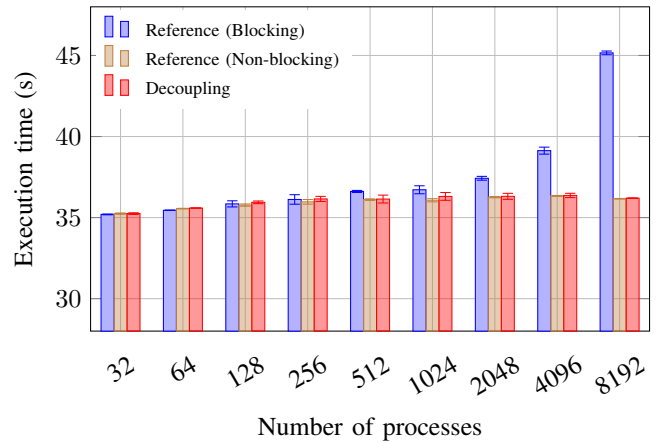

Fig. 6: Weak scaling test of the CG solver using blocking (blue bars), non-blocking (yellow bars) MPI all-to-all operation and the decoupling approach (red bars) on a Cray XC40 supercomputer.

We conduct a weak scaling tests on the Cray XC40 su-percomputer such that each process calculates a subdomain of $120^3$ grid points. Figure 6 reports the average and the standard deviation of execution time. To ensure fair comparison, all implementations run a fixed number of iterations (300) for all tests. The decoupling approach uses one out of every 16 processes for the decoupled operation ($\alpha = 6.25\%$).

---
[2]The CG reference implementations are available at https://htor.inf.ethz.ch/research/nbcoll/cgsolver/

The decoupling approach shows good scalability when the number of processes increases. We observed a near constant execution time in the weak scaling tests from 256 to 8,192 processes. Note that this evaluation uses regular workloads on each process and has tight dependency among operations so that the decoupling approach only has limited benefit from operation overlapping. The performance results show that in such use case, the decoupling model can achieve the same efficiency as the MPI non-blocking operations. When compared to the blocking implementation, the decoupling model performs considerably better, achieving $1.25\times$ improvement on 8,192 processes.

### D. iPIC3D

In this evaluation, we show that the decoupling approach can effectively separate characteristically different operations within one application. We adopt the decoupling approach in a production-quality plasma simulation code, called iPIC3D [18], to improve the performance of two different operations. iPIC3D is a massively parallel Particle-in-Cell code for the simulations of space and astrophysical plasmas. It is used for the preparation of NASA and ESA missions and has a large user base in physics community. iPIC3D is representative for scientific applications running on current petascale supercomputers. The reference implementation is publicly available[3] and has approximately 30,000 LOC.

iPIC3D uses computational particles to sample the statistical distribution of electrons and protons, and solves the electromagnetic field on a computing grid. In production runs, the total number of particles reaches the order of billions and the grid can have millions of grid points. iPIC3D distributes the particles and decomposes the grid over processes. At each iteration, it performs a series of operations such as moments calculation, field solver, particle mover, I/O and data analytics. As the computing grid remains unchanged throughout the simulation, operations on the grid are regular and static. However, the particles can freely move within the global domain, resulting in highly skewed distribution on the processes. Moreover, it is impossible to know a-priori the particle movements. Therefore, operations related to the particles are dynamic and irregular. We decouple two operations related to the particles in this evaluation. We describe the changes in the application and the performance results in the remainder of the section.

*1) Decoupling Particle Communication:* The challenges of the particle communication stem from the fact that each process needs to determine how many particles to receive and from which processes it should receive from. A straightforward solution will have every process to first inform all

[3]iPIC3D is available at https://github.com/CmPA/iPic3D/

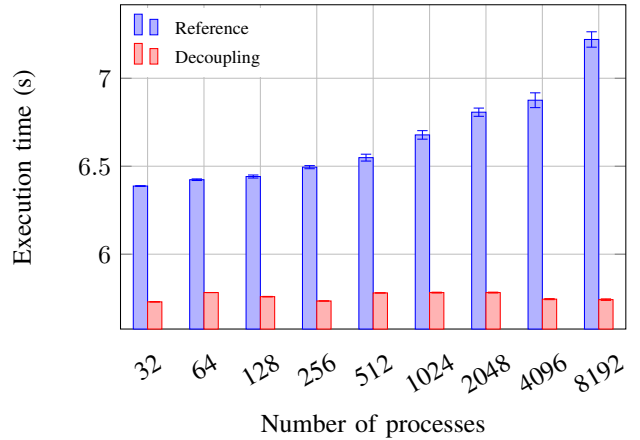

Fig. 7: Weak scaling test of the particle communication in the iPIC3D code using the reference (blue bars) and the decoupled implementations (red bars) on the Cray XC40 testbed.

other processes the amount of particles that it will send and then to transfer the data. On $P$ processes, this process has a complexity of $\mathcal{O}(P^2)$. To further complicate this problem, the pattern of particle communication changes dynamically throughout the execution. Thus, this complexity is paid at each time step.

The reference implementation uses an optimized scheme so that each process only forwards particles to its six direct neighbour processes in the three dimensions. This step is repeated until all particles have reached their destination. This approach reduces the worst scenario to $(Dim_X + Dim_Y + Dim_Z)$ forwarding steps, where $Dim_X$, $Dim_Y$ and $Dim_Z$ are the number of processes in each dimension of the Cartesian communicator. Assume a $10 \times 10 \times 10$ communicator, this scheme has an upper bound of 30 steps, which has a big advantage over the direct solution. It is likely that the particle forwarding can finish before all these steps. Thus after each step, the reference implementation checks the total number of particles for the next step and terminates if there are no more particles to forward.

We note that the operation of communicating particles satisfies several criteria for operations decoupling in Section II-E. First, the operation is highly variable on different processes due to the skewed distribution of particles. Second, the complexity of the operation increases as the number of processes increase. Thus executing the operation on a subset of processes can reduce complexity.

We decouple the operation of particle communication to a separate group of processes, denoted as group $G_1$. The remaining processes form group $G_0$ to perform the calculation of particle trajectories as before. When processes in group $G_0$ are updating the position of particles, if the new position is outside the process subdomain, they stream out the particles to group $G_1$. Processes in group $G_1$ handles the complexity of particle communication internally. They keep processing the arrived particles on a first-come-first-served basis. They aggregate the particles by their destination process and then forward the aggregated particles to the destination process in only one pass. Using the decoupling approach, each particle only requires a maximum of two steps to reach their destination, i.e., from group $G_0$ to $G_1$ and then backward. Execution trace of the reference implementation and the decoupled implementation is presented in Figure 2.

We evaluate our approach using the GEM magnetic reconnection challenge [19]. The experiments use approximately $2 \times 10^9$ particles on 8,192 processes. Figure 7 presents the average and the standard deviation of execution time. The decoupling approach exhibits better scalability compared to the reference implemetation. On 8,192 processes, the decoupling model achieved a $1.3\times$ improvement. Also, the execution time of the decoupling approach remains nearly constant when the system scales up. On the other hand, the reference implementation shows an increased execution time when the number of processes increases. This is consistent with the bound of complexity in the two approaches. Moreover, despite increased variance of particles on a large number of processes, the near constant execution time indicates that the decoupling approach can effectively reduce the impact of imbalance.

*2) Decoupling Particle I/O:* The particles in simulations carry valuable information that requires detailed analysis by physicists. There are three main challenges in saving particle information. First, similar to the particle communication, processes have an irregular and dynamic distribution of particles. Thus the amount of saved data varies on processes and changes at each iteration throughout the simulation. Second, the large amount of particles, in the order of billion or ten billions, typically drains the available memory on processes, making it infeasible to use buffering as a way of optimization (MPI nonblocking operations fall in this category). Finally, collective I/O operations on a large number of processes is intrinsically difficult to optimize, impacted by factors of communication, memory consumption, and file system.

The reference implementation uses MPI I/O operations to save particle information. It supports both the collective I/O (`MPI_File_write_all`) and the non-collective (`MPI_File_write_shared`) for the particle I/O. The MPI collective I/O employs a two-phase optimization to improve performance on large-scale systems. It requires a file view to specify the data layout of each process in the shared file. As the particle distribution on processes changes dynamically, processes need to recalculate their displacement and redefine the file view at each iteration. The non-collective operation does not require to specify the file view but handles the consistence semantics internally inside the MPI library.

The particle I/O operation satisfies several criteria for decoupling, similar to the particle communication. We decouple this operation to a separate group of processes. The remaining processes calculate particle trajectories and stream particles to the I/O group. The I/O group buffers up the arrived particles and save them to the storage using `MPI_File_write_shared`. Note that as the I/O group only performs one operation, it can delicate substantial memory for buffering, reducing the interference with the file system.

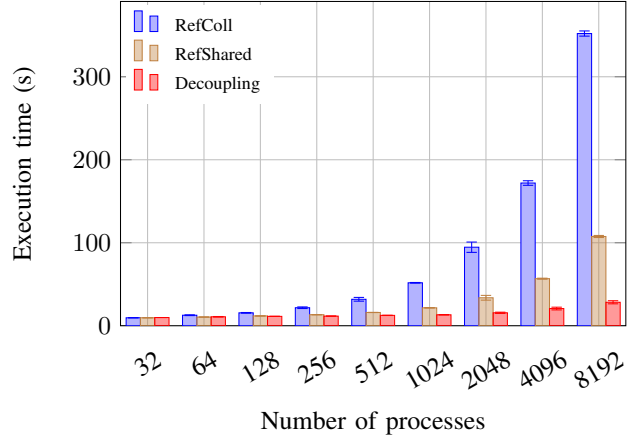

Fig. 8: Weak scaling test of the particle I/O in the iPIC3D code using two reference implementations ( `MPI_File_write_shared` and `MPI_File_write_all` ) and the decoupled implementation on the Cray XC40 testbed.

We use the GEM setup for evaluation and present the results in Figure 8. The decoupling approach uses one out of every 16 processes for the decoupled operation ($\alpha = 6.25\%$). On 8,192 processes, the decoupling strategy achieved $12\times$ and $3\times$ better performance than the two reference implementations, respectively. The decoupling approach started to show performance benefit from 64 processes and shows an increasing improvement over the two reference implementations when the number of processes increases. This result is consistent with the rationale for selecting suitable operations for decoupling. First, the network is more evenly utilized throughout the execution of computation operation, as continuous data flow is generated between two groups. This eliminates the bursty communication pattern as in conventional model. Second, the size of the I/O group is orders smaller than the total number of processes, reducing the complexity of I/O operations. Third, the group has much fewer interaction with the file system because they can support aggressive buffering while the conventional model cannot. Finally, the decoupled operation progresses in parallel with other operations that are performed by the remaining processes, enabling effective pipeline.

## V. RELATED WORK

Many works can effectively address the impact from process imbalance. High-performance implementation of MPI one-sided communication can provide asynchronous OS-bypassing communication [20]. The enclave-based application composition [21] can isolate Operating System/Runtimes in *enclaves*, to address the different requirements from workloads within one compute node. Task-based approaches like Cilk++, Charm++, and Chapel can provide fine-grained execution for better load balancing [7], [8], [9]. Porting production-quality applications into a new programming environment takes significant development efforts and time.

The decoupling technique has shown its effectiveness in developing runtimes on HPC platforms, such as burst-buffer file systems and load balancing framework [22], [13]. However, decoupling general operations at the application-level has not been explored in depth. In this work, we provide a generic programming abstraction and interface for users to adopt the decoupling strategy at the application-level.

Our implementation of the decoupling approach utilizes a stream processing paradigm among groups. The theoretical framework of stream computing stems from Khan process networks (KPN) [23]. KPN provides a simple parallel computing model that connects multiple processes with first-in-first-out channels with flexibility of constructing the process graph. However, the majority of the parallel applications running on current supercomputers are implemented with MPI. Thus it is critical to support this paradigm with minimal modifications in the existing applications. We note that several other approaches can also be used to realize the decoupling mode. For instance, the Active Messages (AM) mechanism supports irregular and fine-grained communication [24]. Another option is Active Pebbles [25], which allows applications to send fine-grained messages and then performs necessary coarsening to enable efficient message passing.

## VI. Conclusions

We proposed a decoupling strategy that separates operations onto groups of processes to improve operation pipelining and to reduce the impact of process imbalance. Our approach utilizes a dataflow processing paradigm among groups. It provides an abstraction and interface for programmers to decouple general operations at application-level. We provide an implementation atop MPI so that existing applications only require incremental changes to adopt the strategy. We evaluated on a petascale supercomputer using scientific and data-analytics applications to show that, with proper operations decoupled and optimized, our approach outperforms traditional implementation and the performance gap increases at scale. We believe that the convenience of adapting existing applications and the considerable performance benefits support the decoupling strategy as an important choice for preparing applications at exascale.

## Acknowledgment

The work was funded by the European Commission through the SAGE project (grant agreement no. 671500). This work was supported by the DOE Office of Science, Advanced Scientific Computing Research, under the ARGO and CENATE projects (award number 66150 and 64386).